# Development of Safety Performance Functions: Incorporating Unobserved Heterogeneity and Functional Form Analysis

**Behram Wali**
Graduate Research Assistant, Department of Civil & Environmental Engineering
The University of Tennessee
311 John D. Tickle Building, 851 Neyland Drive, Knoxville, TN 37909
Tel: 865-306-6677; Fax: 865-974-2503; Email: bwali@vols.utk.edu

**Asad J. Khattak, Ph.D., Corresponding Author**
Beaman Distinguished Professor, Department of Civil & Environmental Engineering
The University of Tennessee
322 John D. Tickle Building, 851 Neyland Drive, Knoxville, TN 37909
Tel: 865-974-7792; Fax: 865-974-2503; Email: akhattak@utk.edu

**Jim Waters, PE.**
Assistant Director, Strategic Transportation Investments Division
Tennessee Department of Transportation
500 Fifth Street, NW, Washington, DC 20001
Tel: 6150-741-2208; Fax:615-532-0353; Email: Jim.Waters@tn.gov

**Deo Chimba, PE, Ph.D.**
Associate Professor, Civil & Architectural Engineering
Tennessee State University
3500 John A Merritt Blvd, Nashville, TN 37209
Tel: 615-963-5430; Email: dchimba@tnstate.edu

**Xiaobing Li**
Graduate Research Assistant, Department of Civil & Environmental Engineering
The University of Tennessee
311 John D. Tickle Building, 851 Neyland Drive, Knoxville, TN 37909
Tel: 865-898-6182; Fax: 865-974-2503; Email: xli77@vols.utk.edu

## ABSTRACT

To improve transportation safety, this study applies Highway Safety Manual (HSM) procedures to roadways while accounting for unobserved heterogeneity and exploring alternative functional forms for Safety Performance Functions (SPFs). Specifically, several functional forms are considered in Poisson and Poisson-gamma modeling frameworks. Using five years (2011-2015) of crash, traffic, and road inventory data for two-way, two-lane roads in Tennessee, fixed- and random-parameter count data models are calibrated. The models account for important methodological concerns of unobserved heterogeneity and omitted variable bias. With a validation dataset, the calibrated and uncalibrated HSM SPFs and eight new Tennessee-specific SPFs are compared for prediction accuracy. The results show that the statewide calibration factor is 2.48, suggesting rural two-lane, two-way road segment crashes are at least 1.48 times greater than what HSM SPF predicts. Significant variation in four different regions in Tennessee is observed with calibration factors ranging between 2.02 and 2.77. Among all the SPFs considered, fully specified Tennessee-specific random parameter Poisson SPF outperformed all competing SPFs in predicting out-of-sample crashes on these road segments. The best-fit random parameter SPF specification for crash frequency includes the following variables: annual average daily traffic, segment length, shoulder width, lane width, speed limit, and the presence of passing lanes. Significant heterogeneity is observed in the effects of traffic exposure-related variables on crash frequency. The study shows how heterogeneity-based models can be specified and used by practitioners for obtaining accurate crash predictions.

## INTRODUCTION & BACKGROUND

As core tools provided in Highway Safety Manual (HSM), published by the American Association of State Highway Officials (AASHTO), Safety Performance Functions (SPFs) are used to estimate expected crash frequencies at a particular facility. Importantly, SPFs that can accurately predict crashes are valuable to state department of transportations (DOTs) as it helps in identifying areas with potential safety concerns. Crash predictive models are used to statistically estimate the expected crash frequency for particular facility type with specified "base" conditions (*1*). The default values of key facility characteristics are used in developing HSM SPFs. If base conditions are not met, HSM provides crash modification factors (CMFs) that can be multiplied with base-case predictions from HSM SPF to capture the differences between base and jurisdiction-specific conditions. Since crash frequency and associated under- and over-dispersion in crash data can vary significantly across jurisdictions, AASHTO clearly recognizes the need for calibrating HSM SPFs to specific jurisdictions (*1*). Compared to uncalibrated counterparts, calibrated HSM predictive models can provide more accurate crash predictions by accounting for variations in crash data and associated factors. However, when enough data is available, HSM allows and encourages transportation agencies to develop jurisdiction- or state-specific SPFs in obtaining more accurate crash predictions.

Calibration of HSM predictive models and development of new SPFs have been performed by several states throughout the U.S. (*2-4*). By using localized data, jurisdiction-specific SPFs are developed by regressing crash frequency as a function of certain explanatory factors. While jurisdiction-specific SPFs (compared with HSM SPFs) can better represent local conditions, traffic crash frequencies and associated factors (e.g. traffic volumes) can still vary significantly across similar, or even identical, road geometry and conditions within a jurisdiction (*5*). The correlations between crashes and associated factors can be heterogeneous, and it is important to correct for heterogeneity in modeled relationships that can arise from a number of observed and unobserved factors relating to (but not limited to) (*5-9*):

- Driver behaviors
- Vehicle types
- Socioeconomic factors
- Traffic and pavement characteristics
- Road geometrics
- Variations in police accident recording thresholds
- Other time and space related unobserved factors.

For a complete review of methodological challenges in crash frequency modeling, interested readers are referred to Lord & Mannering (*10*).

As part of a research project performed by researchers at the University of Tennessee to facilitate the implementation of new HSM procedures in the state of Tennessee, main objectives of this study are to apply HSM predictive models for rural two-lane, two-way roads, compute calibration factors, and explore the need for developing Tennessee-specific SPFs. Before conducting a detailed empirical analysis of rural two-lane, two-way road safety, crash rates are analyzed as an effective *"first brush"* tool to quantify relative safety in different regions within Tennessee. Methodologically, several functional forms based on HSM guidelines and other forms prevalent in literature are considered. Importantly, the development of rigorous SPFs undertaken in this study accounts for unobserved heterogeneity. The models can be used for more accurate prediction of crashes. Finally, out-of-sample prediction forecasts are generated to evaluate the performance of all competing models. Consideration of unobserved heterogeneity in the development of SPFs is critical, as ignoring heterogeneity in the effects of explanatory factors can lead to inconsistent, biased estimates. While this important methodological concern is adequately recognized and extensively addressed in extant safety literature, such methods are rarely used in practice by state DOTs for more accurate prediction of crashes. Notably, the scope of this study is limited to accounting for random heterogeneity as opposed to the possibility of systematic heterogeneity due to non-linearity.

## METHODOLOGY

Present study involves the following tasks: crash rate analysis, calibrating HSM SPF models to better represent local conditions, and development of jurisdiction-specific SPFs. Different techniques employed to achieve the tasks are briefly explained in this section.

## Crash Rates

Crash rate is commonly used for safety evaluation of roadway facilities in practice. In this analysis, crash rates are estimated using two different measures of exposure, i.e., crash rates by vehicles miles travelled (VMT), and crash rates by segment length (*1*). Specifically, crash rates by VMT is calculated using Equation 1 (*1*):

$$R = \frac{C*100,000,000}{V*365*N*L} \quad (1)$$

Where: R = Crash rate per VMT;
C = Total crashes in study period (five years);
V = Traffic volume using Annual Average Daily Volumes (AADT);
N = Number of years of data; and
L = Length of roadway segment in miles.

Likewise, crash rates by segment length is calculated using Equation 2 (*1*):

$$R = \frac{C}{N*L} \quad (2)$$

Where: R = Crash rate per mile of segment;
N = Number of years of data; and
L = Length of roadway segment in miles.

## Calibration Factor Estimation

SPF for rural two-lane, two-way road segments in HSM is developed using data from selected states in the U.S. Equation 3 represents rural two-lane, two-way SPF, and can be applied when certain base conditions (as documented in HSM) meet local jurisdiction conditions to which HSM SPF is applied (*1*).

$$N_{SPF} = AADT * L * 365 * 10^{-6} * e^{-0.312} \quad (3)$$

Where: $N_{SPF}$ = Base predicted number of crashes in study period for site $i$;
$L$ = Length of roadway segment (miles); and
$AADT$ = Annual average daily traffic during study period.

SPF in Equation 3 is estimated for specific base conditions related to roadway geometric features. Whenever base-case conditions are not met, CMFs can be used that account for the differences in local jurisdiction specific geometric features and base conditions assumed in HSM SPF. That is, crashes obtained using Equation 3 can be multiplied with CMFs as:

$$N = N_{SPF} \times CMF_1 \times CMF_2 \times CMF_3 \times \ldots\ldots\ldots. \times CMF_i \quad (4)$$

Where: $N$ = Adjusted predicted crash frequency in local jurisdiction;
$CMF_i$ = Crash modification factor(s) for road segment features from HSM base conditions.

After calculating predicted crash frequency (accounting for non-base jurisdiction specific conditions), the calibration factor can simply be calculated using Equation 5 (*1*):

$$C = \frac{\sum Observed\ Crashes}{\sum Adjusted\ Predicted\ Crashes} = \frac{N_{observed}}{N} \quad (5)$$

Where: $N_{observed}$ = Observed crash frequency in study period; and
$N$ = Adjusted predicted crash frequency in local jurisdiction.

Finally, the calculated calibration factor ($C$) in Equation 5 can be multiplied with HSM SPF (Eq. 3) for predicting rural two-lane, two-way road segment crashes in a specific jurisdiction. The resulting SPF then becomes:

$$N = C * (AADT * L * 365 * 10^{-6} * e^{-0.312}) \quad (6)$$

In what follows, uncalibrated HSM SPF (Eq. 3) will be referred to as Model 1, whereas HSM SPF with calibration (Eq. 6) referred to as Model 2.

## Tennessee-Specific Safety Performance Functions

When enough data is available, it is recommended that users develop jurisdiction-specific SPFs (*1*). In addition to better crash forecasts, developing state-specific SPFs can help in network screening and evaluation of engineering treatments at a site or project level. Given the discrete non-negative data nature of crashes, count data modeling techniques are typically used to model crash frequencies as a function of explanatory variables. Common techniques include Poisson Generalized Linear Models (GLMs) and Negative Binomial GLMs (*10*).

### *Poisson and Negative Binomial Regressions*

For Poisson models, the probability of having a specific number of crashes "$n$" at road segment "$i$" is written in Equation 7 (*10*):

$$P(n_i) = \frac{\exp(-\lambda_i)\lambda_i^n}{n_i!} \quad (7)$$

Where: $P(n_i)$ = probability of crash occurring at segment "$i$", "$n$" times per specific time-period; and
$\lambda_i$ = Poisson parameter for segment "$i$" which is numerically equivalent to segment "$i$" expected crash frequency per year $E(n_i)$.

Formally, $\lambda_i$ can be viewed as a log link function of a set of explanatory factors using Equation 8 (*10*):

$$\ln(\lambda_i) = \beta(X_i) \quad (8)$$

Where: $X_i$ = Vector of explanatory variables; and
β = Vector of estimable parameter estimates.

Poisson function defined in Equation 7 and 8 can be maximized by the standard maximum likelihood procedure (*11*). Application of Poisson regression to over-dispersed crash data can result in inappropriate results. If the mean and variance of crash data are not equal, corrective measures are applied to Equation 8 by adding an independently distributed error term, $\varepsilon$, as:

$$\ln(\lambda_i) = \beta(X_i) + \varepsilon_i \quad (9)$$

Where: $\exp(\varepsilon_i)$ = Gamma-distributed error term with mean one and variance α (*12*).

Following (*13*), if α is statistically significantly different from zero, negative binomial regression should be favored, otherwise the Poisson model is more appropriate.

### *Unobserved Heterogeneity in Crash Frequency Modeling*

As discussed earlier, it is likely that correlations between key explanatory variables and crash frequency may not be consistent across multiple rural two-lane, two-way segments. There are several compelling reasons to expect intrinsic unobserved heterogeneity (*11; 14; 15*). For instance, data used for crash frequency analysis usually have a limited set of variables and other variables influencing crash frequencies are usually not available. Furthermore, if key variables are omitted from analysis and too few variables are included in modeling, parameter estimates may be biased and inaccurate. One way to address this issue is to allow parameter estimates to vary across observations (*15*). As such, random parameters can be included in estimation framework as (*14; 15*):

$$\beta_i = \beta + \varphi_i \quad (10)$$

Where: $\varphi_i$ = randomly distributed term with any pre-specified distribution (e.g. normal distribution) with mean zero and variance $\sigma^2$ (*15*).

With Equation 10, Poisson parameter in Equation 8 becomes:

$$\lambda_i|\varphi_i = EXP(\beta X) \quad (11)$$

And, Poisson parameter in Equation 9 in Poisson-Gamma model becomes:

$$\lambda_i|\varphi_i = EXP(\beta X + \varepsilon_i) \quad (12)$$

Finally, likelihood function for a random-parameter model can be maximized through the maximum simulated likelihood using Equation 13 (*11*):

$$LL = \sum_i ln \int_{\varphi_i}^{i} g(\varphi_i)P(n_i|\varphi_i)d\varphi_i \quad (13)$$

Where: g(.) = Probability density function of randomly distributed term with pre-specified distribution (e.g. normal distribution) with mean zero and variance $\sigma^2$.

Different distributions are tested for random parameters such as normal, log-normal, uniform, triangular, and tent. However, density function based on normal distribution resulted in best fit. For simulation purposes, 200 Halton draws are used. More details on random parameter models can be found in the literature (*10; 15*).

### *Functional Forms*

To develop Tennessee-specific SPFs, different functional forms based on original HSM form and functional forms used by other researchers are considered using both Poisson and Negative Binomial regression techniques (*3; 13*).

**First Form (Models 3 and 4)** Functional forms for Tennessee-specific SPFs in Model 3 and 4 are similar to HSM base SPF (Eq. 3) in terms of simple structure and minimal data requirements. Thus, model 3 (Poisson distribution) and model 4 (negative binomial distribution) are of the form:

$$N_{TN-SPF} = AADT * L * 365 * 10^{-6} * e^{\beta_o} \quad (14)$$

Poisson and negative binomial regressions based on functional form in Equation 14 are equivalent to constant-only regression models with exposure as an offset variable. Exposure is calculated as $AADT * L * 365 * 10^{-6}$ (*1*).

**Second Form (Models 5 and 6)** Second functional form tested is similar to HSM functional form (Eq. 13) in terms of minimal data requirements. For example, only AADT and segment length are used as two potential explanatory variables in Models 5 (Poisson regression) and 6 (negative binomial regression). However, functional form differs from HSM functional form (Equation 13) in terms of variable specification, and is:

$$N_{TN-SPF} = exp(\beta_o) * AADT^{\beta_1} * Seglen^{\beta_2} \quad (15)$$

Where $\beta_o$, $\beta_1$, and $\beta_2$ are parameters to be estimated. Functional form in Equation 15 is equivalent to regressing crash frequency on natural logarithm of AADT and segment length, and is used by several researchers in modeling crash frequencies (*4; 11; 13*).

**Third Form (Models 7 and 8)** Third functional form is using Equation 16 (*10*):

$$\ln(N_{TN-SPF}) = \beta_o + \sum_{i=1}^{P} \beta_i X_i \quad (16)$$

Where: $X_i$ = Matrix of explanatory factors; and

$\beta_i$ = Column-vector of parameter estimates associated with each variable in matrix $X_i$.

Model 7 refers to SPF based on Poisson regression whereas model 8 is based on negative binomial regression.

**Fourth Form (Model 9 and 10)** Finally, Model 9 incorporates unobserved heterogeneity in crash data in addition to all available variables as in Model 7 and 8. This can potentially provide insights into heterogeneous effects of different factors on crash frequency. SPFs in Model 9 and 10 are developed using simulation based random parameter Poisson and negative binomial modeling techniques. The functional form is:

$$N_{TN-SPF} = \exp^{[\beta_o + \sum_{i=1}^{Q} \gamma_i Z_i + \sum_{i=1}^{P} \beta_i X_i]} \quad (17)$$

Where: $Z_i$ = Matrix of explanatory factors of random parameters;

$\gamma_i$ = Column-vector of parameter estimates associated with each variable in matrix $Z_i$;

$X_i$ = Matrix of explanatory factors of fixed parameters; and

$\beta_i$ = Column-vector of parameter estimates associated with each variable in matrix $X_i$.

## Model Validation & Goodness-of-fit Measures

One major objective of estimating more realistic crash frequency models is to enhance crash forecast accuracy. To compare estimated models in terms of model fit and out-of-sample predictions, data was randomly divided into two categories, one for model training and the other for model testing/validation. Specifically, 70% of data is used for model fitting. All models (Model 3-10) are estimated and fitted with training data. The remaining 30% of data (N=90) is used for model testing/validation of HSM SPFs (Model 1-2) and Tennessee-specific SPFs (Model 3-10). To quantify uncertainties in out-of-sample predictions, Mean Absolute Error (MAE) (*12*), Root Mean Square Error (RMSE), and Mean Prediction Bias (MPB) are calculated (*12*). MAE and RMSE give the average magnitude of variability in predictions, i.e., smaller values are preferred. Unlike MAE and RMSE, MPB can be positive or negative. A positive value of MPB indicates SPF is overestimating number of crashes, whereas a negative value implies underestimation. For evaluating goodness-of-fit and statistical adequacy of all models: log-likelihood at convergence, McFadden R-square, Akaike Information Criteria (AIC), and Bayesian Information Criteria (BIC) are used (*12; 16*). A lower value of AIC and BIC indicates a relatively better model.

## Data Assembly

To meet study objectives, significant efforts went into assembling data from different sources manually. Specifically, a statewide crash and roadway inventory database is used to calibrate HSM SPFs and develop Tennessee-specific new rural two-lane, two-way road segments SPFs. Crash, traffic, and roadway geometrics data are collected for a 5-year period (2011-2015). Five years of crash and geometric data are used for both calibration of HSM SPFs and development of Tennessee-specific SPFs. For calibration factor analysis, it is conducted both for each year and an average of 5 years' data. For data collection, inventory data on rural two-way, two-lane roadway segments across Tennessee (with a minimum segment length of 0.10 miles) were collected and compiled from the Tennessee Department of Transportation's (TDOT) Enhanced Tennessee Roadway Information Management System (E-TRIMS) (*17*). A total of 14,777 road segments were identified. Then, a random sample of 299 homogeneous roadway segments with complete data was obtained from original 14,777 sample. The random sample accounts for diverse geographical conditions across Tennessee with 61 segments from Region 1 (Knoxville area), 61 segments from Region 2 (Chattanooga area), 104 segments from Region 3 (Nashville area), and 72 segments from Region 4 (Memphis area). Next, 5-year crash data for the selected roadway segments were extracted manually from E-TRIMS through crash summary reports. To ensure the accuracy of manually extracted data, a computer program was used to assign crashes to each of the 299 roadway segments. Doing so revealed successful and accurate matching of crashes for the sampled 299 roadway segments. As shown in Figure 1, roadway inventory data (e.g., segment length) and traffic data (e.g., AADT and speed limits) were also manually extracted from TDOT's digital image viewer (*17*) and traffic count data program (*18*).

To account for omitted variable bias, data on additional correlates of crashes, as recommended by HSM (*1*), were also collected. Note that data on some of these variables could not be used due to unavailability, including horizontal and vertical alignment, roadside hazard rating, and automated speed enforcement. Notably, the heterogeneity model specifications account for omitted variable biases that may arise due to missing information on important correlates (*5; 11*).

**(PLACE FIGURE 1 ABOUT HERE)**

## RESULTS

### Descriptive Statistics

Table 1 presents descriptive statistics of crash frequencies on rural two-lane, two-way roads in Tennessee, key and additional variables. Based on their distributions, key summary statistics, and extraction from a well-organized and integrated state database, underlying data is of reasonable quality. For example, average 5 years' crashes across 299 rural two-lane, two-way segments are distributed with a mean of 1.247 and standard deviation of 1.883 which highlights slight over-dispersion in crash data. Regarding key variables, the mean of average 5 years' AADT is 1828, whereas mean segment length is observed to be 1.149 miles. Lighting is present on approximately 24% of these roads, whereas center line rumble strips and passing lanes are present on approximately 19% and 27% of roadways respectively. Descriptive statistics of other variables can be interpreted similiarly.

**(PLACE TABLE 1 ABOUT HERE)**

### Crash rates by vehicles miles travelled and segment length

Table 2 provides average crash rates for all four regions and regionwide crash rates by VMT and segment length. Average crash rate per 100-million VMT for all regions is 287.62 with a standard deviation of 384.19. Based on simple crash rates, region 4 appears to be the safest with a crash rate of 239.34 crashes/100-million VMT, followed by Region 3 with 282.93 crashes/100-million VMT.

### Calibration Factors (CF) Results

Two types of calibration factors are calculated, 1) Base Calibration Factors ($CF_{Base}$), and 2) Adjusted Calibration Factors ($CF_{Adj}$). $CF_{Base}$ are estimated by applying HSM SPF (Eq. 3) on AADT and segment length only. $CF_{Adj}$ are estimated by applying Equation 4 in estimating crashes while incorporating CMFs for cases where Tennessee-specific roadway geometrics deviate from HSM default values. The calibration procedure is applied for all regions combined as well as separately for each region. In addition, to account for temporal variations, the calibration procedure is both applied for an average of 5-year data, as well as separately for each year. However, the year-wide calibration factors did not exhibit significant variations.

$CF_{Base}$ and $CF_{Adj}$ for the entire state is found to be 2.980 and 2.489 respectively, showing that rural two-lane, two-way road segment crashes are at least 1.48 times greater than what HSM SPF predicts even after accounting for non-base Tennessee specific conditions. The calibration factor ranged between 2.023 and 2.776 for four regions in Tennessee. For Region 1 and 3, $CF_{Adj}$ is estimated to be 2.584 and 2.776 respectively. For Region 2 and 4, they are 2.444 and 2.023 respectively. Overall, calibration factor results suggest that observed crashes on Tennessee rural two-lane, two-way roads are significantly greater than crashes predicted by calibrated HSM safety performance functions.

**(PLACE TABLE 2 ABOUT HERE)**

## Tennessee-Specific Safety Performance Functions

In this section, results of Tennessee specific SPFs are presented. Specifically, eight different SPFs are developed based on different distributional assumptions (Poisson and Negative Binomial) and different functional forms. All fixed-parameter models (models 3-8) are estimated via standard maximum likelihood procedures, whereas random-parameter models (model 9-10) are estimated via simulated maximum likelihood procedures. All models are derived from a systematic process to include most important variables on basis of statistical significance, specification parsimony, and intuition. First, a series of ordinary least square regressions were estimated to spot correlations and patterns in the data. Next, a series of Poisson and Negative Binomial regressions (both fixed and random parameter) were estimated. Specifically, all variables from Table 1 were tested. Finally, statistically significant variables retain in final model specifications.

### *Model Selection and Performance Comparison*

Before detailed discussion of Tennessee-specific SPFs, goodness-of-fit measures of all estimated models with different distributional assumptions and functional forms are presented in Table 3. Following (*12; 16*), BIC and AIC can be used to evaluate competing nested and non-nested models. Among all estimated models, Model 5 (Poisson SPF based on logarithms of AADT and segment length) has the lowest AIC and BIC indicating relatively superior "in-sample" fit, followed by Model 3 (Poisson SPF based on HSM functional form), and finally, Model 9 (Random Parameter Poisson SPF). Note that goodness-of-fit measures (e.g. log-likelihood at convergence and likelihood-ratio test statistic) presented in Table 3 are indicators for "explanatory" power of competing models. Importantly, "explaining" vs. "predicting" are two different dimensions for which statistical models are estimated (*19*). While AIC is derived from a predictive viewpoint, it is an indicator of "in-sample" fitting capabilities of competing models (*19*), and not an indicator of "out-of-sample" forecasts accuracy. In other words, a model may have high out-of-sample forecast errors with lower AIC.

**(PLACE TABLE 3 ABOUT HERE)**

While SPFs based on only two covariates (e.g. logarithms of AADT and segment length in Model 5-6), performs relatively best, it is interesting to note that Tennessee-specific SPF based on HSM functional form (which only includes estimation of intercept) performs relatively well in fitting "training" data. SPFs based on including all covariates (third form) have the highest AIC and BIC values (lower in-sample fit).

### *Modeling Results*

Table 4 and 5 present results of all Tennessee-specific SPFs for rural two-lane, two-way roads. Specifically, Table 4 summarizes results of fixed parameter SPFs based on Poisson and Negative Binomial distributions, and different functional forms. Whereas, Table 5 summarizes results of random parameter Poisson and Negative Binomial SPFs. Referring to parameter estimates in Table 4 and 5 for different models, a positive sign on parameter estimate shows that specific variable is positively correlated with crash frequency, and vice versa. For instance, in Model 5, AADT and segment length are positively correlated with crash frequency. This finding is in agreement with the extant traffic safety literature (*3; 4; 13*). Likewise, in Model 7, shoulder width is negatively associated with crash frequency. Overall, results shown in Table 4 suggest the absence of over-dispersion, which favors statistical superiority of Poisson regression based SPFs. This finding can also be confirmed in Table 3 where AICs and BICs of Poisson based SPFs are almost equal to AICs and BICs of their negative binomial counterparts.

Coming to results in Table 5, random parameter Poisson and Negative Binomial SPFs resulted in better statistical fit relative to fixed parameter counterparts as in Model 7-8 in Table 4, as is shown by smaller AIC and BIC values in Table 3 as well. Specifically, random parameter estimation technique helps in capturing heterogeneous associations between response outcome and explanatory variables. For instance, in random parameter Poisson regression (Model 9), both AADT and segment length are observed to be normally distributed random parameters, suggesting that these variables' effects vary across different rural two-lane, two-way road segments in Tennessee. It is observed that increase in shoulder width is associated with smaller crash frequency. Contrarily, increase in speed limit, lane width, and presence of passing lanes are associated with increase in crash frequency. Incorporating presence of significant unobserved heterogeneity results in significantly better AIC and BIC values, and McFadden R-square (0.471 for random parameter Poisson compared to 0.375 for fixed parameter Poisson). Also, results in Table 5 suggest that random parameter Poisson regression is statistically superior to random parameter Negative Binomial regression. This is intuitive as negative binomial regression models are typically used to capture over-dispersion in crash data. With random-parameter Poisson regressions, it is likely that majority of over-dispersion is captured in the form of unobserved heterogeneity and thus no (or little) over-dispersion may be left in data for random parameter negative binomial based regression model to capture (*20*).

**(PLACE TABLE 4 ABOUT HERE)**

**(PLACE TABLE 5 ABOUT HERE)**

### *Out of Sample Forecast Evaluation*

While goodness-of-fit statistics presented in Table 5 provide valuable insights regarding "in-sample" fit of all estimated models, "out-of-sample" forecast accuracy cannot be readily inferred from statistics presented in Table 3. As such, a model's true "out-of-sample" forecast capabilities are evaluated by using holdout set technique and calculating statistics that help evaluate "out-of-sample" forecast capabilities of all estimated models. Results of out-of-sample forecast errors of all estimated models are presented in Table 6. Smaller values of MAE and RMSE are desirable. Positive MPB values show that SPF is over-estimating, while negative values show that particular SPF is under-estimating. Among all models tested, Model 9 (TN-Specific Random Parameter Poisson SPF) exhibited the best out-of-sample forecast capabilities with lowest MAE, RMSE, and MPB, followed by Model 5, and Model 3.

**(PLACE TABLE 6 ABOUT HERE)**

Finally, "out-of-sample" mean-estimated over mean-observed number of crashes for HSM SPFs (with and without calibration), Tennessee-specific Poisson SPF with logarithms of AADT and segment length as explanatory variables (Model 5), and random parameter Poisson SPF with all covariates included (Model 9) are shown in Figure 2. Compared to uncalibrated HSM SPF, "out-of-sample" forecasts for calibrated HSM are closer to the mean equivalence line (red line in Figure 2). Also, compared to HSM SPFs (calibrated and uncalibrated) and Tennessee-specific Poisson SPF based on logarithms of AADT and segment length only, the out-of-sample predictions for Tennessee-specific random parameter Poisson SPF are more evenly distributed across the mean-equivalence line (Figure 2). Given all statistical evidence, random parameter based Poisson SPF can be used in practice in Tennessee and can facilitate generation of more accurate crash predictions. The advantage of random parameter SPFs is that more realistic predictions of crashes can be obtained that can better identify Potential for Safety Improvement (PSI) sites, and appropriate countermeasures can be developed. To facilitate application of aforementioned SPFs in engineering practice, Graphical User Interfaces (GUI) are needed.

**(PLACE FIGURE 2 ABOUT HERE)**

## CONCLUSIONS

The main objectives of this study are to apply HSM predictive models for rural two-lane, two-way roads, compute calibration factors, and explore the need for developing Tennessee-specific SPFs. Before conducting detailed empirical analysis of rural two-lane, two-way road safety, crash rates are analyzed as an effective *"first brush"* tool to quantify relative safety in different regions within Tennessee. For development of SPFs, several functional forms based on HSM guidelines and other prevalent in the literature are considered. In particular, unobserved heterogeneity is accounted for by developing fixed- and random-parameter count data models that can be used for more accurate prediction of crashes on Tennessee rural two-lane, two-way roads.

The average crash / 100-million VMT for all regions is 287.62 with a standard deviation of 384.19. Based on regionwide crash rate analysis, Region 4 (Memphis area) appears to have the lowest crash rate at 239.34 crashes /100-million VMT, followed by region 3 (Nashville), with 282.93 crashes /10-million VMT. The calibration of HSM SPFs for rural two-lane, two-way roads revealed that average 5-year calibration factor for all regions is 2.5. This shows that these road segment crashes are at least 1.5 times greater than what HSM SPF predicts after applying calibration factors. However, some differences are observed for regionwide calibration factors as opposed to collective calibration factors for all regions. For instance, while average 5-year calibration factors for Region 1 and 2 are around 2.5, the calibration factors for Region 3 and 4 are different from other regions. Specifically, average 5-year calibration factors for Region 3 and 4 are around 2.8 (highest among all regions) and 2.0 (lowest among all regions) respectively. Overall, calibration factor results suggest observed crashes on Tennessee rural two-lane, two-way roads are significantly greater than crashes predicted by calibrated HSM safety performance functions.

The calibration of HSM SPF to match local conditions of Tennessee improved the prediction accuracies of HSM SPF; especially, out-of-sample forecast errors for calibrated HSM SPF are found to be lower than Tennessee-specific SPFs based on including all variables in model specifications and almost similar to Tennessee-specific SPF based on HSM functional forms. However, Tennessee-specific Poisson SPF based on logarithms of AADT and segment length as only covariates performed better than calibrated HSM SPF and Tennessee-specific SPF based on HSM functional form. Finally, fully specified Tennessee-specific random parameter Poisson SPF outperformed all competing SPFs in forecasting out-of-sample crashes on Tennessee rural two-lane, two-way roads. This provides compelling empirical evidence that unobserved heterogeneity should be accounted for in estimating Tennessee-specific SPFs, and ignoring which can result in biased and inaccurate inferences and forecasts. The methodology provided in this study offers a way to quantify heterogeneous correlations of safety performance (e.g., a relationship between AADT and crash frequency) and identify the PSI for each site. Appropriate countermeasures can be developed for sites with larger magnitudes of parameters, which imply greater PSIs. While the present study accounts for random (unobserved) heterogeneity, systematic heterogeneity (e.g., due to non-linearity) is not comprehensively covered. As future work, both systematic and random heterogeneity components are recommended to be further explored in crash frequency modeling. Without explicitly accounting for both systematic and random heterogeneity, it is impossible to discern the true source of heterogeneity (systematic or random).

## ACKNOWLEDGEMENTS

The authors would like to thank Tennessee Department of Transportation for supporting this research. Special thanks are due to Mr. Steve Allen, Mr. Jeff Murphy, Mr. Zane Pannell, Mr. David A. Duncan for their timely guidance in data collection efforts. The views expressed in this paper are those of the authors. Support was provided by the U.S. DOT funded Southeastern Transportation Center and the U.S. DOT through the Collaborative Sciences Center for Road Safety, a consortium led by The University of North Carolina at Chapel Hill in partnership with The University of Tennessee. The contribution of Ms. Megan Lamon in proof-reading the manuscript is highly appreciated.

## REFERENCES

[1] AASHTO. Highway Safety Manual, Volume 2.In, Washington DC, 2010.
[2] Sun, X., Y. Li, D. Magri, and H. Shirazi. Application of highway safety manual draft chapter: Louisiana experience. *Transportation Research Record: Journal of the Transportation Research Board*, No. 1950, 2006, pp. 55-64.
[3] Brimley, B., M. Saito, and G. Schultz. Calibration of Highway Safety Manual safety performance function: development of new models for rural two-lane two-way highways. *Transportation Research Record: Journal of the Transportation Research Board*, No. 2279, 2012, pp. 82-89.
[4] Mehta, G., and Y. Lou. Calibration and development of safety performance functions for Alabama: Two-lane, two-way rural roads and four-lane divided highways. *Transportation Research Record: Journal of the Transportation Research Board*, No. 2398, 2013, pp. 75-82.
[5] Mannering, F. L., and C. R. Bhat. Analytic methods in accident research: Methodological frontier and future directions. *Analytic Methods in Accident Research*, Vol. 1, 2014, pp. 1-22.

[6] Akbar, M., R. Khan, M. T. Khan, B. Alam, M. Elahi, B. Wali, and A. A. Shah. Methodology for Simulating Heterogeneous Traffic Flow at Intercity Roads in Developing Countries: A Case Study of University Road in Peshawar. *Arabian Journal for Science and Engineering*, 2017, pp. 1-16.
[7] Arvin, R., M. Khademi, and H. Razi-Ardakani. Study on mobile phone use while driving in a sample of Iranian drivers. *International journal of injury control and safety promotion,* Vol. 24, No. 2, 2017, pp. 256-262.
[8] Liu, J., A. J. Khattak, and B. Wali. Do safety performance functions used for predicting crash frequency vary across space? Applying geographically weighted regressions to account for spatial heterogeneity. *Accident Analysis & Prevention,* Vol. 109, 2017, pp. 132-142.
[9] Khattak, A. J., and B. Wali. Analysis of volatility in driving regimes extracted from basic safety messages transmitted between connected vehicles. *Transportation Research Part C: Emerging Technologies,* Vol. 84, 2017, pp. 48-73.
[10] Lord, D., and F. Mannering. The statistical analysis of crash-frequency data: a review and assessment of methodological alternatives. *Transportation Research Part A: Policy and Practice,* Vol. 44, No. 5, 2010, pp. 291-305.
[11] Kamrani, M., B. Wali, and A. J. Khattak. Can Data Generated by Connected Vehicles Enhance Safety? Proactive Approach to Intersection Safety Management. *Transportation Research Record: Journal of the Transportation Research Board (Accepted),* No. 2659, 2017.
[12] Washington, S. P., M. G. Karlaftis, and F. L. Mannering. *Statistical and econometric methods for transportation data analysis*. CRC press, 2010.
[13] Khattak, A. J., A. J. Khattak, and F. M. Council. Effects of work zone presence on injury and non-injury crashes. *Accident Analysis & Prevention,* Vol. 34, No. 1, 2002, pp. 19-29.
[14] Li, X., A. J. Khattak, and B. Wali. Large-Scale Traffic Incident Duration Analysis: The Role of Multi-agency Response and On-Scene Times. *Transportation Research Record: Journal of the Transportation Research Board.,* No. 2616, 2017.
[15] Wali, B., A. Ahmed, and N. Ahmad. An ordered-probit analysis of enforcement of road speed limits.In *Proceedings of the Institution of Civil Engineers-Transport*, Thomas Telford Ltd, 2017. pp. 1-10.
[16] Bozdogan, H. Model selection and Akaike's information criterion (AIC): The general theory and its analytical extensions. *Psychometrika,* Vol. 52, No. 3, 1987, pp. 345-370.
[17] E-TRIMS. https://e-trims.tdot.tn.gov.
[18] TDOT Traffic History. https://www.tdot.tn.gov/APPLICATIONS/traffichistory.
[19] Shmueli, G. To explain or to predict? *Statistical science,* Vol. 25, No. 3, 2010, pp. 289-310.
[20] El-Basyouny, K., and T. Sayed. Accident prediction models with random corridor parameters. *Accident Analysis & Prevention,* Vol. 41, No. 5, 2009, pp. 1118-1123.

**LIST OF TABLES**

**LIST OF FIGURES**

**TABLE 1 Descriptive Statistics of Key Variables**

| | Variable | N | Mean | Std. Dev. | Min | Max |
|---|---|---|---|---|---|---|
| **Crash frequencies** | Total 5 years Crashes | 299 | 6.234 | 9.413 | 0 | 73 |
| | Average 5 years crashes | 299 | 1.247 | 1.883 | 0 | 14.6 |
| | Average 5 years crashes (rounded to nearest integer) | 299 | 1.230 | 1.951 | 0 | 15 |
| **Key variables** | Total 5 years AADT | 299 | 9141.0 | 11453.0 | 302 | 73056 |
| | Average of 5 years AADT | 299 | 1828.0 | 2290.0 | 60.4 | 14611.2 |
| | Segment length (miles) | 299 | 1.149 | 1.303 | 0.1 | 7.2 |
| **Additional variables** | Lane width (feet) | 299 | 10.436 | 1.258 | 7 | 12 |
| | Combined shoulder width (feet) | 299 | 3.348 | 2.593 | 0 | 12 |
| | Gravel shoulder | 299 | 0.291 | 0.455 | 0 | 1 |
| | Asphalt Concrete shoulder | 299 | 0.421 | 0.495 | 0 | 1 |
| | Turf shoulder | 299 | 0.288 | 0.453 | 0 | 1 |
| | Lighting present | 299 | 0.241 | 0.428 | 0 | 1 |
| | Speed limit (miles per hour) | 299 | 39.866 | 9.562 | 20 | 55 |
| | Presence of C/L rumble strips | 299 | 0.187 | 0.399 | 0 | 1 |
| | Presence of passing lane | 299 | 0.268 | 0.443 | 0 | 1 |
| | Presence of short-four-lane-section | 299 | 0.010 | 0.100 | 0 | 1 |
| | Presence of two way left turn lane | 299 | 0.017 | 0.128 | 0 | 1 |

**TABLE 2 Crash Rates by VMT and Segment Length**

| Crash Rates/100-million VMT | | | | | |
|---|---|---|---|---|---|
| Area | N | Mean | Std. Dev. | Min | Max* |
| **All Regions** | 299 | 287.62 | 384.19 | 0 | 3320.88 |
| **Region 1** | 61 | 331.93 | 390.97 | 0 | 2265.16 |
| **Region 2** | 61 | 309.09 | 533.22 | 0 | 3320.88 |
| **Region 3** | 104 | 282.93 | 331.28 | 0 | 1570.97 |
| **Region 4** | 73 | 239.34 | 291.81 | 0 | 1331.77 |
| Crashes per each mile of roadway per year | | | | | |
| Area | N | Mean | Std. Dev. | Min | Max |
| **All Regions** | 299 | 1.759 | 3.793 | 0 | 33.99 |
| **Region 1** | 61 | 2.100 | 4.817 | 0 | 31.43 |
| **Region 2** | 61 | 1.351 | 1.646 | 0 | 7.00 |
| **Region 3** | 104 | 2.247 | 4.885 | 0 | 33.99 |
| **Region 4** | 73 | 1.121 | 1.681 | 0 | 9.82 |

Notes: (*) the extremely high maximum crash rates (as indicated by “max” column) are for segments with low number of crashes but also very short segments i.e., $\sim 0.15\ miles$); N is sample size; Std. Dev. is standard deviation.

**TABLE 3 Goodness-of-fit Statistics of TN-Based SPFs**

| Functional Form* | TN-Specific SPFs | N | LL(Null) | LL (Convergence) | DF | AIC | BIC |
|---|---|---|---|---|---|---|---|
| **First Form*** | Model 3 | 209 | -233.022 | -233.022 | 1 | 468.04 | 471.39 |
| | Model 4 | 209 | -232.39 | -232.39 | 2 | 468.79 | 475.48 |
| **Second Form**** | Model 5 | 209 | -410.68 | -227.4 | 3 | 460.80 | 470.83 |
| | Model 6 | 209 | -325.37 | -227.39 | 4 | 462.78 | 476.15 |
| **Third Form***** | Model 7 | 209 | -410.68 | -256.03 | 7 | 526.06 | 549.46 |
| | Model 8 | 209 | -325.37 | -261.03 | 8 | 527.31 | 551.60 |
| **Fourth Form****** | Model 9 | 209 | -466.91 | -246.6 | 9 | 509.31 | 541.10 |
| | Model 10 | 209 | -466.9135 | -246.608 | 10 | 513.22 | 546.64 |

Notes: N is the sample size; LL(Null) is log-likelihood of constant only model; LL(Convergence) is log-likelihood at convergence.; DF is number of parameters estimated; AIC is Akaike Information Criteria; and BIC is Bayesian Information Criteria.

1. (*) First form is $N_{TN-SPF} = AADT * L * 365 * 10^{-6} * e^{\beta_o}$;
2. (**) Second form is $N_{TN-SPF} = exp(\beta_o) * AADT^{\beta_1} * Seglen^{\beta_2}$;
3. (***) Third form is $\ln(N_{TN-SPF}) = \beta_o + \sum_{i=1}^{P} \beta_i X_i$;
4. (****) Fourth form is $N_{TN-SPF} = \exp[\beta_o + \sum_{i=1}^{Q} \gamma_i Z_i + \sum_{i=1}^{P} \beta_i X_i]$

**TABLE 4: Estimation Results (Tennessee Specific Fixed Parameter Count Data Models) (Models 3-8)**

| | Model 3[a] | | Model 4[b] | | Model 5[a] | | Model 6[b] | | Model 7[a] | | Model 8[b] | |
|---|---|---|---|---|---|---|---|---|---|---|---|---|
| | β | t-stat | β | t-stat | β | t-stat | β | t-stat | β | t-stat | β | t-stat |
| Constant | 0.7468 | 12.36 | 0.774 | 10.93 | -5.456 | -10.89 | -5.462 | -10.78 | -1.765 | -4.99 | -1.949 | -5.05 |
| Exposure*(offset) | 1 | --- | 1 | --- | --- | --- | --- | --- | --- | --- | --- | --- |
| AADT (ln form) | --- | --- | --- | --- | 0.783 | 12.17 | 0.784 | 12.03 | --- | --- | --- | --- |
| Segment length (ln form) | --- | --- | --- | --- | 0.904 | 14.35 | 0.904 | 14.26 | --- | --- | --- | --- |
| AADT (in thousands) | --- | --- | --- | --- | --- | --- | --- | --- | 0.271 | 10.41 | 0.291 | 7.822 |
| Segment length | --- | --- | --- | --- | --- | --- | --- | --- | 0.456 | 12.41 | 0.511 | 10.23 |
| Shoulder width | --- | --- | --- | --- | --- | --- | --- | --- | -0.165 | -4.73 | -0.172 | -4.21 |
| Speed limit | --- | --- | --- | --- | --- | --- | --- | --- | 0.024 | 2.81 | 0.023 | 1.21 |
| LW dummy (1 if lane width >= 10) | --- | --- | --- | --- | --- | --- | --- | --- | 0.075 | 0.37 | 0.169 | 0.73 |
| Passing lane dummy | --- | --- | --- | --- | --- | --- | --- | --- | 0.203 | 1.29 | 0.225 | 1.17 |
| Over-dispersion | --- | --- | 0.0515 | 0.856 | --- | --- | 0.005 | 0.119 | --- | --- | 0.191 | 2.3 |
| **Summary Statistics**** | | | | | | | | | | | | |
| McFadden-$R^2$ | 0 | | 0 | | 0.446 | | 0.301 | | 0.375 | | 0.025 | |
| $\chi^2$ Statistic[c] | --- | | --- | | 366.56 | | 195.96 | | 309.29 | | 151.89 | |
| Prob > Critical $\chi^2$ | --- | | --- | | 0.0000 | | 0.0000 | | 0.0000 | | 0.0000 | |
| N | 209 | | 209 | | 209 | | 209 | | 209 | | 209 | |

Notes: (*) Exposure is AADT*Segment length*365*0.000001; (**) Other in-sample goodness-of-fit statistics are provided in Table 3; β is the parameter estimate; (a) is SPF based on Poisson distribution; (b) is SPF based on negative-binomial distribution; (c) No Chi-square statistic is applicable for Model 3 and 4 as they are constant-only models where all other coefficients are forced to be zero; (---) means Not-Applicable.

**TABLE 5: Estimation Results (Tennessee Specific Random Parameter Count Data Models) (Models 9 and 10)**

| | Model 9 [a] | | | Model 10 [b] | | |
|---|---|---|---|---|---|---|
| **Variables** | β | SE | t-stat | β | SE | t-stat |
| ***Fixed Parameters*** | | | | | | |
| Constant | -1.9 | 0.314 | -6.04 | -1.9 | 0.321 | -5.91 |
| Shoulder width | -0.166 | 0.033 | -5.02 | -0.166 | 0.033 | -4.96 |
| Speed limit | 0.016 | 0.007 | 2.01 | 0.018 | 0.009 | 1.97 |
| LW dummy (1 if lane width >= 10, 0 otherwise) | 0.335 | 0.106 | 3.16 | 0.321 | 0.202 | 1.58 |
| Passing lane dummy | 0.266 | 0.131 | 2.031 | 0.264 | 0.149 | 1.789 |
| ***Random Parameters*** | | | | | | |
| AADT (in thousands) | 0.289 | 0.027 | 10.52 | 0.278 | 0.025 | 11.12 |
| *standard deviation* | *0.036* | *0.015* | *2.37* | *0.035* | *0.011* | *3.181* |
| Segment length | 0.543 | 0.043 | 12.56 | 0.541 | 0.041 | 13.19512 |
| *standard deviation* | *0.16* | *0.025* | *6.2* | *0.17* | *0.026* | *6.53846* |
| Over-dispersion | --- | --- | --- | *0.003* | *0.07* | *0.052* |
| **Summary Statistics*** | | | | | | |
| McFadden-$R^2$ | 0.471 | | | 0.469 | | |
| $\chi^2$ Statistic | 440.60 | | | 438.12 | | |
| Prob > Critical $\chi^2$ | 0.0000 | | | 0.0000 | | |
| N | 209 | | | 209 | | |

Notes: (*) Other in-sample goodness-of-fit statistics are provided in Table 3; β is parameter estimate; SE is standard error; (a) is random-parameter SPF based on Poisson distribution; (b) is random-parameter SPF based on negative-binomial distribution; (---) means Not-Applicable.

**TABLE 6: Out-of-Sample Forecast Evaluation**

| Functional Forms | Model | Mean Absolute Error | Root Mean Square Error | Mean Prediction Bias |
|---|---|---|---|---|
| HSM SPFs | HSM-SPF (Model 1) | 2.009 | 1.417 | -0.710 |
| | (HSM-SPF)* $CF_{Adj}$ (Model 2) | 1.615 | 1.270 | -0.049 |
| First Form | $N_{TN-SPF} = AADT * L * 365 * 10^{-6} * e^{\beta_o}$ | | | |
| | **TN-SPF- Poisson Regression (Model 3)** | **1.569** | **1.252** | **0.18** |
| | TN-SPF- NB Regression (Model 4) | 1.606 | 1.267 | 0.179 |
| Second Form | $N_{TN-SPF} = exp(\beta_o) * AADT^{\beta_1} * Seglen^{\beta_2}$ | | | |
| | **TN-SPF- Poisson Regression (Model 5)** | **1.265** | **1.125** | **0.051** |
| | TN-SPF- NB Regression (Model 6) | 1.266 | 1.125 | 0.05 |
| Third Form | $\ln(N_{TN-SPF}) = \beta_o + \sum_{i=1}^{P} \beta_i X_i$ | | | |
| | TN-SPF- Poisson SPF (Model 7) | 2.208 | 1.485 | 0.354 |
| | TN-SPF- NB SPF (Model 8) | 3.327 | 1.824 | 0.515 |
| Fourth Form | $N_{TN-SPF} = \exp[\beta_o + \sum_{i=1}^{Q} \gamma_i Z_i + \sum_{i=1}^{P} \beta_i X_i$ | | | |
| | **TN-Specific Random Parameter Poisson SPF (Model 9)** | **1.199** | **1.095** | **0.034** |
| | TN-Specific Random Parameter Negative Binomial SPF (Model 10) | 1.224 | 1.315 | 0.041 |

Note: Highlighted in bold are models that perform better in terms of out-of-sample predictions.

**Response Variables:**
1. Total crashes
2. Total Injury crashes

**Key Explanatory Variables:**
1. AADT
2. Segment length

**Additional Explanatory Variables:**
1. Lane width
2. Shoulder type
3. Combined shoulder width
4. Presence or absence of centerline rumble strips
5. Presence or absence of passing lane
6. Presence or absence of short four lane section
7. Presence or absence of two way left turn lane
8. Presence or absence of roadway lighting

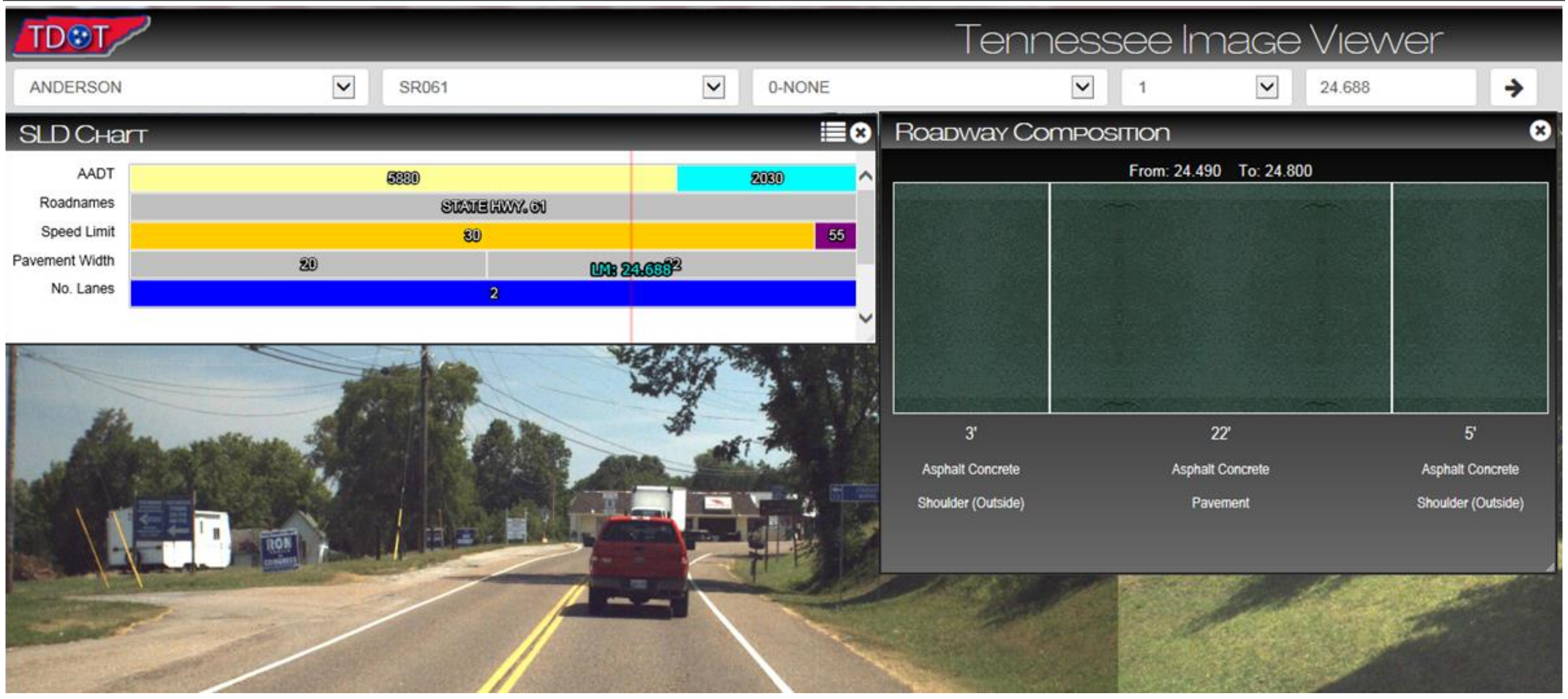

**FIGURE 1 Illustration of E-TRIMS and Tennessee Image Viewer software for manual extraction of key roadway geometric data.**

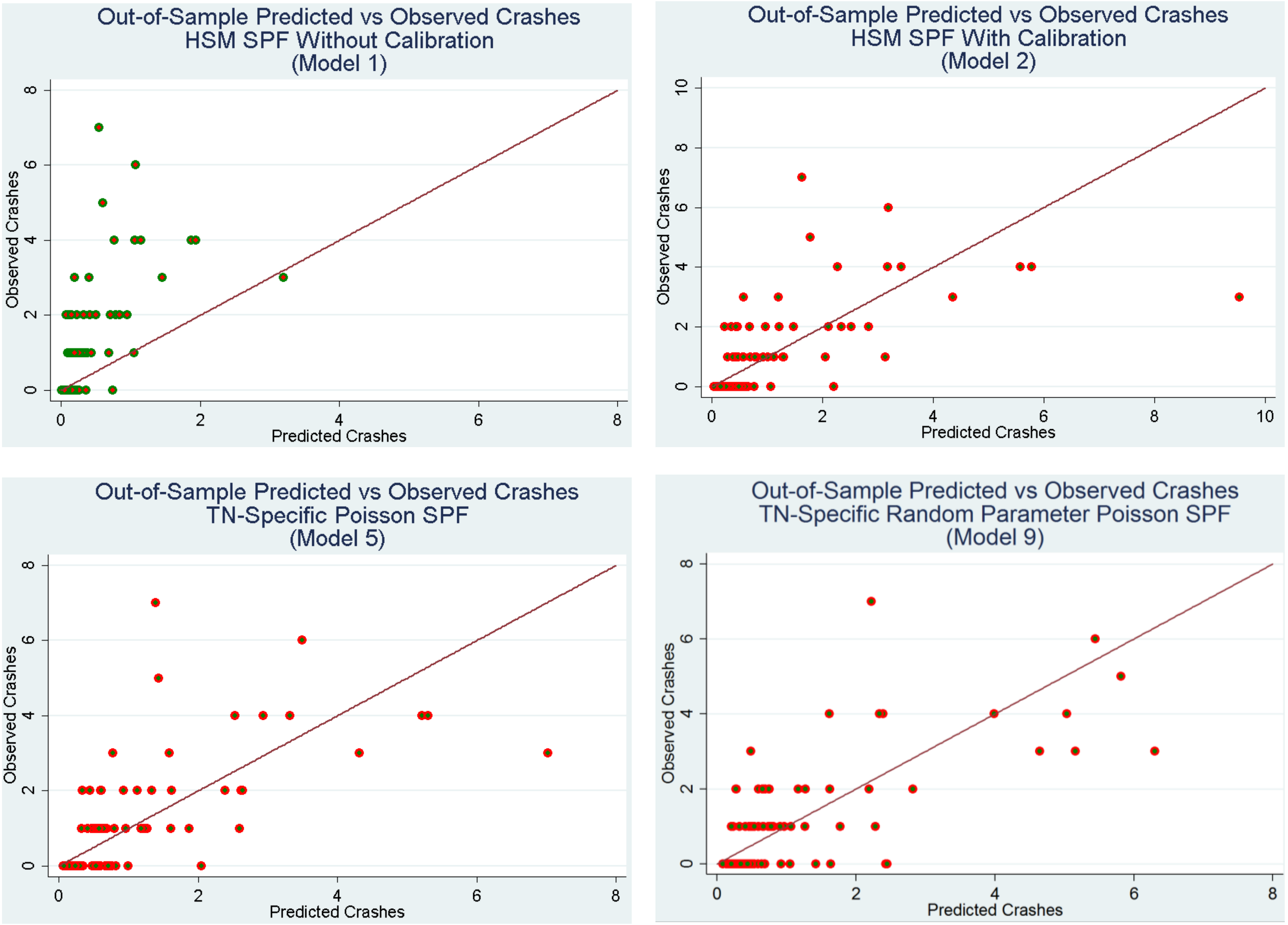

**FIGURE 2 Out-of-Sample predicted vs. observed crashes for HSM SPF (with and without calibration) and TN-Specific SPFs.**
Note: Based on fixed and random parameter Poisson SPFs (Red line indicates the equivalence of mean-estimated and mean-observed values).